\documentclass[12pt]{article}

\usepackage{latexsym}

\usepackage{graphicx}

\begin{document}

%\begin{flushright}
%Sofia University\\
%\end{flushright}
%%%%%%%%%%%%%%%%%%%%%%%%%%%%%%%%%%%%%%%%%%%%%%%%%%%%%%%%%%%%%%%%%%%

\title{Asymptotically and non-asymptotically flat static black rings in charged dilaton gravity }

\author{
     Stoytcho S. Yazadjiev \thanks{E-mail: yazad@phys.uni-sofia.bg}\\
{\footnotesize  Department of Theoretical Physics,
                Faculty of Physics, Sofia University,}\\
{\footnotesize  5 James Bourchier Boulevard, Sofia~1164, Bulgaria }\\
{\footnotesize and }\\
{\footnotesize  Bogoliubov Laboratory of Theoretical Physics,
                JINR, 141980 Dubna, Russia}\\
}

\date{}

\maketitle

\begin{abstract}
We systematically derive the asymptotically flat five dimensional black rings in EMd gravity
by using the sigma model structure of the dimensionally reduced field equations. New non-asymptotically
flat EMd black ring solutions in five dimensions  are also constructed and their physical properties are analyzed.
\end{abstract}

%%%%%%%%%%%%%%%%%%%%%%%%%%%%%%%%%%%%%%%%%%%%%%%%%%%%%%%%%%%%%%%%%%%

%\draft
\sloppy

\section{Introduction}

An interesting development in the black holes studies is the discovery of the black ring solutions
of the five dimensional Einstein equations by Emparan and Reall \cite{ER1}, \cite{ER2}. These are asymptotically
flat solutions with an event horizon of topology $S^2\times S^1$ rather the much more familiar $S^3$ topology.
Since the Emparan and Reall's discovery many explicit examples of black ring solutions were found in various gravity theories \cite{E}-\cite{P}. Elvang was able to apply Hassan-Sen transformation to the solution \cite{ER2} to find a charged black ring in the bosonic sector of the truncated heterotic string theory\cite{E}. A supersymmetric black ring in five dimensional minimal supergravity was derived in \cite{EEMR1} and then generalized to the case of concentric rings in \cite{GG1} and \cite{GG2}. A static black ring solution of the five dimensional Einstein-Maxwell gravity was found by Ida and Uchida in \cite{IU}. In \cite{EMP} Emparan derived "dipole black rings" in Einstein-Maxwell-dilaton (EMd) theory
in five dimensions. Static and asymptotically flat black rings solutions in five dimensional EMd gravity with
arbitrary dilaton coupling parameter $\alpha$ were presented in \cite{KL} without giving any derivation.

In this paper we systematically derive asymptotically flat EMd black ring solutions. Moreover, we also derive
new non-asymptotically flat EMd black ring solutions and analyze their properties. Our motivation is the following.
First, such solutions are interesting in their own right. Second, despite of the discovery of the explicit
examples of black EMd rings, the systematic construction of new higher dimensional solutions in EMd gravity
has not been accomplished in comparison with the four dimensional case. In particular, solutions so far found are
not so many.

In order to achieve our goals
we first consider the $D$-dimensional  EMd theory in static spacetimes. After performing
dimensional reduction
along the timelike Killing vector we show that the norm of the Killing vector and
the electric potential parameterize  a $GL(2,R)/SO(1,1)$ sigma model coupled to
$(D-1)$-dimensional  euclidean gravity. Then, we show that the $GL(2,R)$
subgroup that preserves the asymptotic flatness is $SO(1,1)$. Applying the $SO(1,1)$
transformation to the five dimensional static neutral black ring solution we obtain
the static EMd black rings.  The non-asymptotically flat EMd black rings can be
obtained by acting on the neutral black rings with special elements of $SL(2,R)\subset GL(2,R)$.

\section{General equations and  solution construction}

The EMd gravity in $D$-dimensional spacetimes is described by the action\footnote{In what follows we consider  theories with $\alpha\ne 0$.}

\begin{equation}
S= {1\over 16\pi} \int d^Dx \sqrt{-g}\left(R - 2g^{\mu\nu}\partial_{\mu}\varphi \partial_{\nu}\varphi  -
e^{-2\alpha\varphi}F^{\mu\nu}F_{\mu\nu} \right).
\end{equation}

The field equations derived from the action are

\begin{eqnarray}
R_{\mu\nu} &=& 2\partial_{\mu}\varphi \partial_{\nu}\varphi + 2e^{-2\alpha\varphi} \left[F_{\mu\rho}F_{\nu}^{\rho} - {g_{\mu\nu}\over 2(D-2)} F_{\beta\rho} F^{\beta\rho}\right], \\
\nabla_{\mu}\nabla^{\mu}\varphi &=& -{\alpha\over 2} e^{-2\alpha\varphi} F_{\nu\rho}F^{\nu\rho}, \\
&\nabla_{\mu}&\left[e^{-2\alpha\varphi} F^{\mu\nu} \right]  = 0 .
\end{eqnarray}

We consider static spacetimes and denote the Killing vector by $\xi$ ($\xi={\partial\over \partial t}$).
The metric of the static spacetime can be written in the form

\begin{equation}
ds^2 = - e^{2U}dt^2 + e^{-{2U\over D-3 }} h_{ij}dx^idx^j
\end{equation}

where $U$ and $h_{ij}$ are independent of the time coordinate $t$.
The staticity of the dilaton and electromagnetic fields imply

\begin{equation}
{\cal L}_{\xi}\varphi = 0, \,\,\,  {\cal L}_{\xi} F = 0 .
\end{equation}

Since ${\cal L}_{\xi} F= di_{\xi}F$ there exist locally an electromagnetic potential $\Phi$ such that

\begin{equation}
F = e^{-2U}\xi \wedge d\Phi.
\end{equation}

 In terms of the potentials
$U$, $\Phi$, $\varphi$ and $(D-1)$-dimensional metric $h_{ij}$ the field equations become

\begin{eqnarray}
{\cal D}_{i}{\cal D}^{i}U &=& 2{D-3\over D-2 }e^{-2U-2\alpha\varphi}h^{ij}{\cal D}_{i}\Phi {\cal D}_{j}\Phi , \\
{\cal D}_{i}{\cal D}^{i}\varphi &=& \alpha e^{-2U-2\alpha\varphi}h^{ij}{\cal D}_{i}\Phi {\cal D}_{j}\Phi ,\\
&{\cal D}_{i}&\left( e^{-2U-2\alpha\varphi}h^{ij}{\cal D}_{j}\Phi \right) = 0 ,\\
{\cal R}(h)_{ij} &=& {D-2\over D-3} {\cal D}_{i}U {\cal D}_{j}U + 2{\cal D}_{i}\varphi {\cal D}_{j}\varphi
- 2e^{-2\alpha\varphi -2U}{\cal D}_{i}\Phi {\cal D}_{j}\Phi ,
\end{eqnarray}

where ${\cal D}_{i}$ and ${\cal R}(h)_{ij}$ are the covariant derivative and the Ricci tensor with respect to the metric $h_{ij}$. These equations can be derived from the action

\begin{equation}
S = \int d^{(D-1)}x \sqrt{h} \left[{\cal R}(h) - {D-2\over D-3}h^{ij}{\cal D}_{i}U {\cal D}_{j}U -
2h^{ij}{\cal D}_{i}\varphi {\cal D}_{j}\varphi    + 2e^{-2\alpha\varphi - 2U}h^{ij}{\cal D}_{i}\Phi {\cal D}_{j}\Phi\right].
\end{equation}

Let us introduce the symmetric matrix

\begin{eqnarray}
P = e^{(\alpha_{D} -1)U} e^{- (\alpha_{D} + 1)\varphi_{D}}\left(%
\begin{array}{cc}
  e^{2U + 2\alpha_{D}\varphi_{D}} - (1 + \alpha^2_{D})\Phi^2_{D} & - \sqrt{1 + \alpha^2_{D}}\Phi_{D} \\
- \sqrt{1 + \alpha^2_{D}}\Phi_{D}  &  -1 \\\end{array}%
\right)
\end{eqnarray}

where

\begin{equation}
\alpha_{D}= \sqrt{{D-2\over 2(D-3)}} \alpha, \,\,\, \varphi_{D} = \sqrt{{2(D-3)\over (D-2)}} \varphi, \,\,\, \Phi_{D}= \sqrt{{2(D-3)\over (D-2)}} \Phi .
\end{equation}

Then the action can be written in the form

\begin{equation}
S = \int d^{(D-1)}x \sqrt{h} \left[{\cal R}(h) - {1\over 2(1+ \alpha_{D}^2) }{(D-2)\over (D-3)} h^{ij} Sp \left({\cal D}_{i}P {\cal D}_{i}P^{-1} \right)  \right].
\end{equation}

The action is invariant under the symmetry transformations

\begin{equation}
P \longrightarrow GPG^{T}
\end{equation}

where $G\in GL(2,R)$. The matrix $P$ parameterized a coset $GL(2,R)/SO(1,1)$
where $SO(1,1)$ is the stationary subgroup (see below).

\section{Asymptotically flat solutions}

In this section we will restrict ourselves to asymptotically flat  solutions with
\begin{equation}
U(\infty)=0 ,\,\,\, \varphi(\infty)=0 , \,\,\, \Phi(\infty)=0
\end{equation}

which corresponds to

\begin{equation}
 P(\infty)  = \sigma_{3}
\end{equation}

where $\sigma_{3}$ is the third Pauli matrix. The $G(2,R)$ transformations
which preserves the asymptotics satisfy

\begin{equation}
B\sigma_{3}B^{T} = \sigma_{3}.
\end{equation}

Therefore we conclude that  $B\in SO(1,1)$. We parameterize the $SO(1,1)$ group  in the standard way

\begin{equation}
B = \left(%
\begin{array}{cc}
  \cosh(\gamma) & \sinh(\gamma) \\
  \sinh(\gamma) &\cosh(\gamma) \\\end{array}%
\right) .
\end{equation}

Let us consider a static asymptotically flat  solution of $D$-dimensional Einstein equations

\begin{equation}
ds_{0}^2 = -e^{2U_{0}}dt^2 + e^{-{2U_{0}\over D-3}}h_{ij}dx^idx^j
\end{equation}

which is encoded into the matrix
\begin{equation} P_{0}= e^{(\alpha_{D} -1)U_{0} }
\left(%
\begin{array}{cc}
  e^{2U_{0}} & 0 \\
0 & -1 \\\end{array}%
\right) .
\end{equation}

and the metric $h_{ij}$. The $SO(1,1)$ transformations then generate a solution of the $D$-dimensional EMd gravity given by the matrix

\begin{equation}
P = BP_{0}B^{T}.
\end{equation}

and with the same metric $h_{ij}$.

In more explicit form we have

\begin{eqnarray}
e^{U} &=& {e^{U_{0}}\over \left[\cosh^2(\gamma) - e^{2U_{0}}\sinh^2(\gamma) \right]^{1\over 1 + \alpha^2_{D} } },\\
e^{-\varphi_{D}} &=&  \left[\cosh^2(\gamma) - e^{2U_{0}}\sinh^2(\gamma) \right]^{\alpha_{D}\over 1 + \alpha^2_{D} },\\
\Phi_{D} &=& {\tanh(\gamma) \over \sqrt{1 + \alpha^2_{D}} } {1 - e^{2U_{0}}\over 1 - e^{2U_{0}}\tanh^2(\gamma) }.
\end{eqnarray}

Let us restrict the considerations to five dimensional spacetimes. One of the most interesting solutions of the five dimensional  Einstein equations is the black ring solution given by the metric

\begin{eqnarray}\label{BRS}
ds^2_{0} &=& - {F(x)\over F(y)}dt^2 \\ &+& {1\over A^2(x-y)^2} \left[ F(x)(y^2-1)d\psi^2 + {F(x)F(y)\over y^2 -1}dy^2
 + {F^2(y)\over 1-x^2 }dx^2 + F^2(y){1-x^2\over F(x) }d\phi^2\right] \nonumber
\end{eqnarray}

where $F(x)= 1 - \mu x$, $A>0$ and $0<\mu < 1$. The coordinate $x$ is in the range $-1\le x\le 1$ and the
coordinate $y$ is in the range $y\le -1$. The topology of the horizon is $S^2\times S^1$ parameterized by
$(x,\phi)$ and $\psi$, respectively. In order to avoid a conical singularity at $y=-1$ one must demand that
the period of $\psi$ satisfies $\Delta \psi=2\pi\sqrt{1 + \mu}$. If one demands regularity at $x=-1$ the period of
$\phi$ must be $\Delta \phi=2\pi \sqrt{1 +\mu}$. In this case the solution is asymptotically flat and the ring is sitting on the rim of disk shaped membrane with a negative deficit angle. To enforce regularity at $x=1$ one must take $\Delta \phi = 2\pi \sqrt{1-\mu}$ and the solution describes a black ring sitting on the rim of disk shaped hole in an infinitely extended deficit membrane with positive deficit. More detailed analysis of the black ring solution can be found in \cite{ER1}.

The $SO(1,1)$ transformations generate the following EMd solution

\begin{eqnarray}
ds^2 = - \left[\cosh^2(\gamma) - \sinh^2(\gamma){F(x)\over F(y)} \right]^{-2\over 1 + \alpha^2_{5}}{F(x)\over F(y)}dt^2 \\ + {\left[\cosh^2(\gamma) - \sinh^2(\gamma){F(x)\over F(y)} \right]^{1\over 1 + \alpha^2_{5}}\over A^2(x-y)^2} \left[ F(x)(y^2-1)d\psi^2  \right. \nonumber \\ \left. +  {F(x)F(y)\over y^2 -1}dy^2
 + {F^2(y)\over 1-x^2 }dx^2 + F^2(y){1-x^2\over F(x) }d\phi^2\right] ,\nonumber \\
e^{-\varphi_{5}} = \left[\cosh^2(\gamma) - \sinh^2(\gamma){F(x)\over F(y)} \right]^{\alpha_{5}\over 1 + \alpha^2_{5}} , \\
\Phi_{5} = {\tanh(\gamma)\over \sqrt{1 + \alpha^2_{5}} } {1 - {F(x)\over F(y) }\over 1 - \tanh^2(\gamma) {F(x)\over F(y)} } .
\end{eqnarray}

This solution was presented first in \cite{KL} without derivation. Here we have systematically derived it from the
symmetries of  dimensionally reduced EMd equations. Detailed analysis of the solution is given in \cite{KL}. Here we
present only some basic results. The physical properties naturally share many of the same features of the neutral
black ring solution, in particular, the horizon topology is $S^2\times S^1$.  The area of the horizon is

\begin{equation}
{\cal S}_{h\pm} = 8\pi^2  \cosh^{3\over 1 + \alpha^2_{5}}(\gamma) {\mu^2 \sqrt{(1+ \mu)(1 \pm \mu)}\over A^3}
\end{equation}

where the sign $\pm$ corresponds to taking the conical singularity at $x=\pm 1$.  The solution is asymptotically flat and this can be seen by change
of the coordinates

\begin{equation} \label{COOR}
\rho_{1}= {(1+\mu)\sqrt{y^2-1}\over A(x-y)}, \,\,\, \rho_{2}={(1+\mu)\sqrt{1-x^2}\over A(x-y)}, \,\,\, {\tilde \psi}=
{\psi\over \sqrt{1+\mu}} ,\,\,\, {\tilde \phi}=
{\phi\over \sqrt{1+\mu}} .
\end{equation}

Defining $\rho= \sqrt{\rho^2_{1} + \rho^2_{2}}$ and taking the limit $\rho \to \infty$ we obtain

\begin{equation}
ds^2 \sim - dt^2 + d\rho^2_{1} + d\rho^2_{1} + \rho^2_{1} d{\tilde \psi}^2 + \rho^2_{2} d{\tilde \phi}^2
\end{equation}

i.e. the solution is asymptotically flat. Note, however, that if the conical singularity lies at $x=-1$
the asymptotic metric is a deficit membrane. The mass is given by

\begin{equation}
M_{\pm} = {3\pi  \over 4} {\mu \sqrt{(1+\mu)(1\pm \mu)}\over A^2} \left(1 + {2\over 1 + \alpha^2_{5} }\sinh^{2}(\gamma)  \right)
\end{equation}

where the sing $\pm$ corresponds to the location of the conical singularity. The charge is found to be

\begin{equation}
Q_{\pm} = \pi \sqrt{{3\over 1 + \alpha^2_{5}}} \sinh(\gamma)\cosh(\gamma) { \mu\sqrt{(1+\mu)(1\pm \mu)} \over A^2 }.
\end{equation}

The temperature can be found in the standard way by Eucledeanizing the metric:

\begin{equation}
T = {A\over 4\pi\mu \cosh^{3\over 1 + \alpha^2_{5}}(\gamma)}.
\end{equation}

A Smarr-type relation is also satisfied

\begin{equation}
M_{\pm} = {3\over 8}T{\cal S}_{h\pm}  + \Phi_{h}Q_{\pm}
\end{equation}

where $\Phi_{h}$ is the electric potential evaluated at the horizon.

\section{Non-asymptotically flat solutions}

In order to generate non-asymptotically flat solutions we shall consider the matrix

\begin{equation}
N = \left(%
\begin{array}{cc}
 0 & - a^{-1} \\
 a & a \\\end{array}%
\right) \in SL(2,R)\subset GL(2,R) .
\end{equation}

Then we obtain the following EMd solution presented by the matrix

\begin{equation}
P = N P_{0} N^{T}
\end{equation}

i.e.

\begin{eqnarray}
e^{U} &=& {e^{U_{0}}\over \left[a^2(1 - e^{2U_{0}}) \right]^{1\over 1 + \alpha^2_{D} } }, \\
e^{-\varphi_{D}} &=& \left[a^2(1 - e^{2U_{0}}) \right]^{\alpha_{D}\over 1 + \alpha^2_{D} } ,\\
\Phi_{D} &=& - {a^{-2} \over \sqrt{1 + \alpha^2_{D} } } {1\over 1 - e^{2U_{0}} } .
\end{eqnarray}

Applying this transformation to the neutral black ring solution (\ref{BRS})
we obtain the following EMd solution

\begin{eqnarray}
ds^2 &=& - \left[a^2\left(1 - {F(x)\over F(y) }\right) \right]^{{-2\over 1 + \alpha^2_{5}}} {F(x)\over F(y)}dt^2 \\
&+& {\left[a^2\left(1 - {F(x)\over F(y) }\right) \right]^{{1\over 1 + \alpha^2_{5}}} \over A^2 (x-y)^2 }
\left[ F(x)(y^2-1)d\psi^2 + {F(x)F(y)\over y^2 -1}dy^2
 + {F^2(y)\over 1-x^2 }dx^2 + F^2(y){1-x^2\over F(x) }d\phi^2\right] \nonumber \\
e^{-\varphi_{5}} &=& \left[a^2\left(1 - {F(x)\over F(y) }\right) \right]^{\alpha_{5}\over 1 + \alpha^2_{5} } ,\\
\Phi_{5} &=& - {a^{-2} \over \sqrt{1 + \alpha^2_{5} } } {1\over 1 - {F(x)\over F(y)} } .
\end{eqnarray}

\subsection{Analysis of the solution}

From the explicit form of the solution it is clear that there is a horizon at $y=-\infty$. In the near-horizon limit
the metric of the $t-y$ plane is

\begin{equation}
ds^2_{ty}  \approx  a^{2\over 1 + \alpha^2_{5}} F(x) \left[- {dt^2\over a^{6\over 1 + \alpha^2_{5}}\mu |y|} + {\mu \over A^2 |y|^3 }dy^2 \right]
\end{equation}

After performing a coordinate transformation $y = - {4\mu \over A^2 Y^2 }$ we obtain the metric

\begin{equation}
ds^2_{ty} \approx a^{2\over 1 + \alpha^2_{5}} F(x) \left[ - {A^2 Y^2 \over 4 a^{6\over 1 + \alpha^2_{5}} \mu^2}dt^2 + dY^2 \right]
\end{equation}

which is conformal to that of the Rindler space with an acceleration parameter $\omega = {A\over 2 \mu}a^{-3\over 1 + \alpha^2_{5}}$ (note that $F(x)>0$). Further, performing the coordinate transformation $X=Y\cosh(\omega t)$ and $T=Y\sinh(\omega t)$ the metric becomes manifestly conformally flat

\begin{equation}
ds^2_{ty} \approx a^{2\over 1 + \alpha^2_{5}} F(x) \left[- dT^2 + dX^2 \right] .
\end{equation}

This shows that $ty$ metric has  a nonsingular horizon at $y=-\infty$. It is clear that the other terms can also be
continuously extended and we obtain nonsingular near-horizon geometry

\begin{equation}
ds^2 \approx a^{2\over 1 + \alpha^2_{5}} F(x) \left[- dT^2 + dX^2  + {d\psi^2\over A^2 } \right] +
a^{2\over 1 + \alpha^2_{5}} {\mu^2 \over A^2} \left[{dx^2 \over 1-x^2}  + {1-x^2\over F(x)}d\phi^2  \right].
\end{equation}

The constant-time slices through the horizon have metric

\begin{equation}
ds_{h}^2 ={a^{2\over 1 + \alpha^2_{5} }\over A^2} \left[F(x)d\psi^2 + \mu^2\left({dx^2 \over 1-x^2}  + {1-x^2\over F(x)}d\phi^2  \right) \right].
\end{equation}

Further, the analysis of the near-horizon geometry is quite similar to that for the near-horizon geometry of
the neutral black ring solution. As it is seen the both near-horizon geometries are conformal with a constant conformal factor. In order for the metric to be regular at $x=-1$ the period of $\phi$ must be $\Delta \phi=2\pi \sqrt{1+ \mu}$. The regularity at $x=1$ requires $\Delta \phi=2\pi \sqrt{1-\mu}$. Therefore, it is not possible for
the metric to be regular at both $x=-1$ and $x=1$. The regularity at $x=-1$ means that there is a conical singularity
at $x=1$ and vice versa. The deficit angle is

\begin{equation}
\delta_{\pm 1} = 2\pi \left(1 - \sqrt{ 1 \pm  \mu \over 1 \mp \mu } \right).
\end{equation}

The $x\phi$ part of the metric describes a two-dimensional surface with $S^2$ topology  and with a
conical singularity at one of the poles.

In order to analyze the $y\psi$ part of the metric in the limit $y\to -1$, as in the neutral case, we set
$y = - \cosh(\eta/\sqrt{1+\mu})$. We then find that the $y\psi$ part is conformal to

\begin{equation}
ds^2_{y\psi} \approx d\eta^2 + {\eta^2\over 1 +\mu}d\psi^2.
\end{equation}

In order for the metric to be regular at $\eta=0$ the coordinate $\psi$ must be identified with a period
$\Delta \psi= 2\pi \sqrt{1+\mu}$.

The above analysis show that the topology of the horizon is $S^2\times S^1$. The area of the horizon is

\begin{equation}
{\cal S}_{h\pm} = 8\pi^2 \left( {a^{1\over 1+\alpha^2_{5} }\over A }\right)^3 \mu^2 \sqrt{(1+\mu)(1\pm \mu)}
\end{equation}

where the sign $\pm$ corresponds to taking the conical singularity at $x=\pm 1$.

The asymptotic infinity corresponds to $x=y=-1$. One can show that near $x=y=-1$ the solution behaves as

\begin{eqnarray}
ds^2 &\approx& - \left[{A^2 \over 2a^2\mu(1+\mu) } (\rho_{1}^2 + \rho_{2}^2) \right]^{2\over 1+\alpha^2_{5} } dt^2
\\ &+& \left[{A^2 \over 2a^2\mu(1+\mu) } (\rho_{1}^2 + \rho_{2}^2) \right]^{-1\over 1+\alpha^2_{5} } \left[d\rho^2_{1} +
 d\rho^2_{2} + \rho^2_{1}d{\tilde \psi}^2  + \rho^2_{2}d{\tilde \phi}^2  \right] ,\nonumber \\
e^{2\alpha\varphi} &=& \left[{A^2 \over 2a^2\mu(1+\mu) } (\rho_{1}^2 + \rho_{2}^2) \right]^{2\alpha^2_{5}\over 1+\alpha^2_{5} } ,\\
\Phi &=&- {\sqrt{3}\over 2 \sqrt{1+ \alpha^2_{5}} } \left[{A^2 \over 2a^2\mu(1+\mu) } (\rho_{1}^2 + \rho_{2}^2) \right] .
\end{eqnarray}

where the coordinates $\rho_{1}$,  $\rho_{2}$, ${\tilde \psi}$ and ${\tilde \phi}$ are defined in (\ref{COOR}).
It is clear that the solution is not asymptotically flat.

In order to compute the mass of the non-asymptotically solution we need more sophisticated techniques than in the
asymptotically flat case. We use a naturally modified five dimensional version of the  quasilocal formalism  in four dimensions \cite{BY}. Here we will not discuss the quasilocal formalism but present the final result for the mass:

\begin{equation}
M_{\pm} = {\alpha^2_{5} \over 1 + \alpha^2_{5} } {3\pi\mu\sqrt{(1+\mu)(1\pm \mu)}\over 4A^2 } .
\end{equation}

The electric charge is defined by

\begin{equation}
Q= {1\over 8\pi} \int_{\infty} e^{-2\alpha\varphi} F^{\mu\nu}d\Sigma_{\mu\nu}
\end{equation}

and we find

\begin{equation}
Q_{\pm} = -{\sqrt{3}\pi \over \sqrt{1 + \alpha^2_{5}} } {a^2 \mu\sqrt{(1+\mu)(1\pm \mu)}\over A^2} .
\end{equation}

The temperature can be found in the standard way by Euclideanizing the near-horizon metric ($t=-i\tau$) and periodically identifying $\tau$ to avoid new conical singularities.  After doing that
we obtain

\begin{equation}
T = {A\over 4\pi\mu a^{3\over 1 + \alpha_{5}^2} } .
\end{equation}

The electric potential is defined up to an arbitrary additive constant. In the asymptotically flat case
there is a preferred gauge in which $\Phi(\infty)=0$. In the non-asymptotically flat case the electric potential diverges at spacial infinity and there is no preferred gauge. The arbitrary constant, however, can be fixed so that
the Smarr-type relation to be satisfied:

\begin{equation}
M_{\pm} = {3\over 8} TS_{h\pm} + \Phi_{h}Q_{\pm}.
\end{equation}

\section{Conclusion}

In the present paper we considered $D$-dimensional EMd gravity in static spacetimes.
Performing dimensional reduction along the timelike Killing vector we obtain $GL(2,R)/SO(1,1)$ sigma model
coupled to $(D-1)$-dimensional euclidean gravity. Applying $GL(2,R)$ transformations to the five dimensional neutral static black ring solution we were able to find both asymptotically and non-asymptotically flat five dimensional
dilatonic black rings. Non-asymptotically flat dilatonic black rings solutions were analyzed. These solutions suffer from the presence of conical singularities as in the case of the neutral static rings. In the neutral case the
rotation removes the conical singularities. That is why we expect that the rotating non-asymptotically flat EMd black rings will be free from conical singularities. The construction of non-asymptotically flat EMd solutions with rotation is now in progress and the results will be presented elsewhere.

\section*{Acknowledgements}

This work was partially supported by the Bulgarian National Science Fund under Grant MU 408. The author would like
to thank the Bogoliubov Laboratory of Theoretical Physics (JINR) for their kind hospitality.

\end{document}